\documentclass[aps,twocolumn]{revtex4}
\usepackage{graphicx,epsfig}
\usepackage{bbm,bm}
\usepackage{multirow}

\usepackage{ifthen}
\ifthenelse{\isundefined{\texdir}}{\def\texdir{}}{}

\def\be{\begin{equation}}
\def\ee{\end{equation}}
\def\baray{\begin{eqnarray}}
\def\earay{\end{eqnarray}}
\def\2pi{\left(2\pi\right)}

\def\dper{{d_\perp}}

\def\tic{$\surd$}
\def\Dbar{$\overline{\text{D}}$}

\renewcommand{\bar}[1]{\overline{#1}}

\begin{document}

\title{The Production, Spectrum and Evolution of Cosmic Strings in Brane
  Inflation}
\author{Nicholas T. Jones}
\email{nick.jones@cornell.edu}
\author{Horace Stoica}
\email{fhs3@mail.lns.cornell.edu}
\author{S.-H. Henry Tye}
\email{tye@mail.lns.cornell.edu}

\affiliation{Laboratory for Elementary Particle Physics, Cornell
  University, Ithaca, NY 14853
}

\date{\today}

\begin{abstract}
  Brane inflation in superstring theory predicts that cosmic strings 
  (but not domain walls or monopoles) are produced towards the end 
  of the   inflationary epoch. Here, we discuss the production, the
  spectrum and the evolution of such cosmic strings, properties
  that differentiate them from those coming from an abelian Higgs
  model. As D-branes in extra dimensions, some type of cosmic 
  strings will dissolve rapidly in spacetime, while the stable 
  ones appear with a spectrum of cosmic string tensions. Moreover, 
  the presence of the extra dimensions reduces the interaction 
  rate of the cosmic strings in some scenarios, resulting 
  in an order of magnitude enhancement of the number/energy density 
  of the cosmic string network when compared to the field theory case.
\end{abstract}

\maketitle
\section{Introduction}

The cosmic microwave background (CMB) data \cite{cobe,new} strongly
 supports the inflationary universe scenario \cite{guth} to be the
explanation of the origin of the big bang. However, the origin of
the inflaton and its potential is not well understood---a paradigm
in search of a model.

Recently, the brane world scenario suggested by superstring theory
was proposed, where the standard model of the strong and electroweak
interactions are open string (brane) modes while the graviton and the
radions are closed string (bulk) modes.
In a generic brane world scenario, there are three types of light 
scalar modes : (1) bulk modes like radions (\emph{i.e.}~the
sizes/shape of the compactified dimensions) and the dilaton
(\emph{i.e.}~the coupling), (2) brane positions (or relative positions)
and (3) tachyonic modes which are present on non-BPS branes or branes
that are not BPS relative to each other \cite{Sen:1999mg}.  In
general, the bulk modes have gravitational strength couplings (so too
weak to reheat the universe at the end of inflation) and so are not
good inflaton candidates. Neither are the tachyonic modes, which roll
down the potential too fast for inflation. This leaves the relative
brane positions (\emph{i.e.}~brane separation) as candidates for
inflation.  So, natural in the brane world is the brane inflation
scenario \cite{Dvali:1998pa}, in which the inflaton is an open string mode
identified with an inter-brane separation, while the inflaton
potential emerges from the exchange of closed string modes between
branes; the latter is the dual of the one-loop partition function of
the open string spectrum, a property well-studied in string theory 
This interaction is gravitational strength,
resulting in a very weak (that is, relatively flat) potential, ideally
tailored for inflation.

The scenario is simplest when the radion and the dilaton (bulk) modes
are assumed to be stabilized by some unknown non-perturbative bulk
dynamics at the onset of inflation.  Since the inflaton is a brane
mode, and the inflaton potential is dictated by the brane mode
spectrum, it is reasonable to assume that the inflaton potential is
insensitive to the details of the bulk dynamics.

Brane inflation has been shown to be very robust (see 
\emph{e.g.}~\cite{Burgess:2001fx,Garcia-Bellido:2001ky,Jones:2002cv,
Quevedo:2002xw}).
The inflaton potential is essentially dictated by the gravitational
attractive (and the Ramond-Ramond) interaction between
branes.  As the branes move towards each other, slow-roll inflation 
takes place. This yields an almost scale-invariant power spectrum for
the density perturbation. As they reach a distance around the string 
scale, the inflaton potential becomes relatively steep so that the 
slow-roll condition breaks down. Inflation ends when branes collide 
and heat the universe \cite{stw}, which is the origin of the big bang.
Towards the end of the brane inflationary epoch in the brane world,
tachyon fields appear. As a tachyon rolls down its potential,
defects are formed (see \emph{e.g.}~\cite{Alexander:2001ks}). Due to
properties of the superstring theory and the cosmological conditions, 
only cosmic strings (but not domain walls or monopoles) are
copiously produced during the brane collision \cite{Jones:2002cv,
Sarangi:2002yt}. These cosmic strings are Dp-branes with (p-1)
dimensions compactified.  The CMB radiation
data fixes the superstring scale to be close to the grand
unified (GUT) scale, which then determines the cosmic string tensions,
which turn out to have values that are compatible with today's
observation, but may be tested in the near future.

In field theory, one may also devise an abelian Higgs like model 
around the GUT scale to produce cosmic strings towards the end of 
inflation in which the cosmic string tension is essentially a free 
parameter. Although such a model may not be as well motivated as 
brane inflation, it is a possibility, so we aim to find signatures
that distinguish cosmic strings in brane inflation from those coming
from an abelian Higgs model.

In this paper, we explore more closely the production of 
cosmic strings after inflation, the properties of the cosmic strings, 
in particular their tensions and stability, and finally their evolution
to an eventual network. In summary, we find that the final outcome 
depends crucially on the quantitative details of a particular 
brane inflationary scenario being contemplated. 
In some scenarios, the cosmic strings produced 
via the Kibble mechanism may dissolve quickly. It is likely that 
their dissolution (which can happen soon after (re)heating) leads to 
the thermal production of lower-dimensional branes as cosmic strings. 
This is very likely if the (re)heating process is efficient \cite{stw},
since the (re)heat temperature is comparable to the superstring scale. 
In other scenarios, they will evolve to a cosmic string network. 
In this case,
the general properties of the resulting cosmic string network is 
likely to be quite different from that arising from field theory.
The cosmic strings appear as defects of the tachyon condensation and
can be D1 branes or Dp-branes wrapping a (p-1)-dimensional compact
manifold. They yield a spectrum of cosmic string tensions
including Kaluza-Klein modes. Moreover, due to the presence of
the compactified dimensions the interaction rate of the cosmic
strings in some scenarios decreases, and when compared to the case in
ordinary field theory, the result is an increase by orders of
magnitude in the number density of the cosmic string network in our
universe.

\section{A Variety of Brane Inflationary Scenarios}

The brane inflationary scenarios we are interested in have
the string scale close to the GUT scale so we consider only brane
world models which are supersymmetric (post-inflation) at the GUT 
scale. (Supersymmetry is expected to be broken at the TeV scale, 
which is negligible for the physics we are interested in here.)
In the 10-dimensional
superstring theory, the cosmic strings in our 4-dimensional spacetime
shall be D-branes with one spatial dimension lying along the 3 large
spatial dimensions representing our universe. Hence we seek to
enumerate the possible stable configurations of branes of different
dimensionality in 10 dimensions, compactified on a six manifold.  To
be specific, let us consider a typical Type IIB orientifold model 
compactified on $(T^2\times T^2\times T^2)/\mathbbm{Z}_N$ or some 
of its variations (see for instance \cite{Gimon:1996rq}).
The model has $\mathcal N=1$ spacetime supersymmetry.  Although we
shall focus the discussion on Type IIB orientifolds, the underlying
picture is clearly more general.  We seek to categorize stable
configurations of branes which remain after inflation, and give rise
to stable cosmic strings in the universe.  By ``stable'' we mean that
some fraction of the cosmic strings produced is required to persist until
at least the epoch of big-bang nucleosynthesis in order for observable
effects to be generated.

In supersymmetric Type IIB string theory with branes and orientifold
planes, it is well known that only odd (spacial) dimensional
branes are stable.  The conditions for stable
brane configurations are simple given the  compactification manifold;
branes must differ by only 0, 4 or 8 in dimension, and branes of the
same dimension can be angled at right angles in two orthogonal
directions \cite{Berkooz:1996km, Lifschytz:1996iq}. In the generic 
case where the second homotopy class of the compactification manifold 
is $\pi_2 = \mathbbm Z^3$, branes will be stable when wrapping 
2-cycles in the compact manifold.  From these conditions, we formulate
Table~\ref{stable_table} of branes from which we shall build models of
post-inflation cosmology.
\begin{center}\begin{table}[h]
  \begin{tabular}{|l*{5}{|c}|l}
    \cline{2-6}
    \multicolumn{1}{c}{}&\multicolumn{5}{|c|}{Dimension}&\\
    \cline{1-6}
    \small{Stable Branes}&01&23&45&67&89\\\cline{1-6}
    D9&\tic&\tic&\tic&\tic&\tic&
    \multirow{4}{2cm}{\begin{tabular}{lp{2cm}}
	$\left.\phantom{\llap{$\biggl\}^{\bigl\}}$}}\right\}$&
	$\mathbbm R^{3,1}$ branes
    \end{tabular}}\\
    D5$_1$&\tic&\tic&\tic&&\\
    D5$_2$&\tic&\tic&&\tic&\\
    D5$_3$&\tic&\tic&&&\tic\\\cline{1-6}
    D5$_{1,2}$&\tic&&\tic&\tic&&
    \multirow{4}{2cm}{\begin{tabular}{lp{2cm}}
	$\left.\phantom{\llap{$\biggl\}^{\bigl\}}$}}\right\}$&
	cosmic strings
    \end{tabular}}\\
    D5$_{1,3}$&\tic&&\tic&&\tic\\
    D5$_{2,3}$&\tic&&&\tic&\tic\\
    D1$_0$&\tic&&&&\\
    \cline{1-6}
  \end{tabular}
  \caption{Stable configurations of D-branes. The labels on the the 
    D-branes indicate which of the three 2-cycles they wrap in the
    compactification dimensions and an empty spot indicates no
    wrapping/presence.  For simplicity the cosmic strings are placed
    along the 1-direction.}\label{stable_table}
\end{table}\end{center}

In cosmological situations, branes which are non-BPS relative to the
others can be present.  Generally the non-BPS configurations will
decay, and the decay products of many are well known.  For instance a
Dp-D(p-2) brane combination will form a bound state of a Dp
brane with an appropriate amount of ``magnetic'' flux \cite{Gava:1997jt}.
This process is best understood as the delocalisation or ``smearing
out'' of the D(p-2) brane within the Dp brane. This process in the
Dp-D(p-2) brane system is described by the presence of a tachyon field,
an open string that stretches between them. This tachyon
condenses as the D(p-2) brane decays and
leads to a singular ``magnetic'' flux on the Dp brane; this ``magnetic''
flux then spreads out across the Dp brane and diminishes, leaving
the total flux conserved.  In an uncompactified theory, the residual
``magnetic'' field strength then vanishes. 
Since the tachyon in the Dp-D(p-2) brane combination is a complex 
scalar field inside the D(p-2) brane world volume, its 
rolling/condensation allows the formation of D(p-4)-branes as defects.
(The actual formation/production of D(p-4)-branes may require the 
dissolution of a D(p-2)-\Dbar(p-2) pair inside the Dp-brane.)

Another important set of non-BPS brane configurations which will be
generated in early universe brane-world cosmology are branes of the
same dimension oriented at general angles, which will also decay into
branes with magnetic flux, as described above.  There are also special
cases of non-BPS configurations which will not decay; between a D3$_3$
and D5$_1$ brane (or its T-dual equivalents, for instance a D1 and D7
brane) there is a repulsive force as seen in the total interbrane
potential, which includes all gravitational and RR forces, between a
Dp and a Dp$^\prime$-brane (p$^\prime <$p) (in terms of the separation
distance $r$ when $r\gg M_s^{-1}$)
\be
\label{vpot}
V(r) \sim -\frac{4-(p-p^\prime + 2a)}{r^{p-7+a}},
\ee
where $a$ is the number of directions in which the branes are
orthogonal \cite{Lifschytz:1996iq}.  This potential also makes clear
that there is no force between the BPS configurations of branes
described above - those which differ in dimension by 4 and those of
the same dimension which are angled in two orthogonal directions.

Brane world models of inflation require brane anti-brane pairs 
(or branes oriented at non-BPS angles) \cite{Burgess:2001fx,
Garcia-Bellido:2001ky,Jones:2002cv,Quevedo:2002xw}; the inflaton
field is described by the separation between the branes, and its
potential can be organized to give slow roll inflation.  To describe
the Standard Model, we demand a chiral post-inflation brane-world,
which requires that the branes which form our universe are angled in
some dimension; sets of D5$_1$ and D5$_2$ branes will give a stable
chiral low energy effective theory, for instance.

After the compactification to 4-dimensional spacetime, the Planck mass
$M_P= (8 \pi G)^{-1/2} = 2.4 \times 10^{18}$ GeV is given by
\be
\label{mp}
g_s^2 M_P^2 = M_s^2(M_sr_1)^2(M_sr_2)^2(M_sr_3)^2/\pi
\ee 
where $M_s$ is the superstring scale and
the compactification volumes (of the (45)-, (67)- and (89)-directions)
are $V_i=l_i^2 = (2 \pi r_i)^2$ for $i=$ 1, 2 and 3 respectively.
Here, $M_sr_i \ge $1.
The string coupling $g_s$ should be large enough for 
non-perturbative dynamics to stabilize the radion and the dilaton 
modes (but not too large that a dual version of the model has a weak
coupling). We expect the string coupling generically to be
$g_s {\ \lower-1.2pt\vbox{\hbox{\rlap{$>$}\lower5pt\vbox{\hbox{$\sim$}}}}\
}1$. 
To obtain a theory with a weakly coupled sector in the low energy
effective field theory (\emph{i.e.}~the standard model of strong and
electroweak 
interactions with weak gauge coupling constant), it then seems
necessary to have the brane world picture \cite{Kakushadze:1998wp}.
Suppose the D5$_1$-branes contain the standard model open string 
modes, then
\be
\label{gutg}
g_s \simeq \alpha_{\text{GUT}} (M_sr_1)^2
\ee
where $\alpha_{\text{GUT}} \simeq 1/25$ is the standard model coupling at 
the GUT scale, which is close to the superstring scale $M_s$. 
This implies that $(M_sr_1)^2 \sim$ 30. 
If some standard model modes come from 
D5$_2$-branes, or from open strings stretching between D5$_1$- and 
D5$_2$-branes, then $(M_sr_2)^2 \sim$ 30.

In the early universe, additional branes (and anti-branes) may be
present. Additional branes must come in pairs of brane-anti-brane
(or at angles), so that the total (conserved) RR charge in the 
compactified volume remains zero. Any even dimensional D-branes 
are non-BPS and so decay rapidly. 
The Hubble constant during inflation is roughly
\be 
\label{hubblec}
H^2 \simeq M_s^4/M_P^2
\ee

In Table~\ref{inflatons}, we
catalogue the various brane anti-brane pairs (provided that they 
are separated far enough apart) which can inflate the 4
dimensional Minkowski brane world volume of D5$_1$- and D5$_2$-branes. 
Towards the end of inflation, a tachyon field appears and its
rolling allows the production of defects.  \emph{A priori}, the 
defects (only cosmic strings here) which are allowable under the 
rules of K-theory \cite{Witten:1998cd} may be produced immediately 
after inflation, when the tachyon field starts rolling down.  
Following Eq.(\ref{mp}) and Eq.(\ref{hubblec}), we see that the 
Hubble size $1/H$ during this epoch is
much bigger than any of the compactification radii,
\be
  H^{-1} \gg r_i
\ee
This means that the Kibble mechanism is capable of producing only 
defects with vortex winding in the three large spacial dimensions.
The cosmological production of these defects towards the end of 
inflation are referred to as ``Kibble''  
in Table~\ref{inflatons}. During this epoch, the universe is 
essentially cold and so no thermal production of any defect is 
possible. Generically, codimension-one non-BPS defects may also
be produced. However, these decay rapidly and will be ignored here.
 
\begin{center}\begin{table}[h]
    \begin{tabular}{|l|*{4}{c|}}\cline{2-5}
      \multicolumn{1}{c|}{}&Inflation&
      \multicolumn{3}{|c|}{Cosmic String Types}\\
      \cline{1-1}\cline{3-5}
      Inflaton& Possible& Allowed&
      Kibble& Thermal\\\hline
      D$(9-\bar9)$&$\times$&&&\\\hline
      D$(7-\bar7)_{1,2}$
      &\tic&1$_0$,3$_1$,3$_2$,5$_{1,2}$&5$_{1,2}$&-\\\hline
      D$(7-\bar7)_{1,3}$
      &\tic&1$_0$,3$_1$,3$_3$,5$_{1,3}$&5$_{1,3}$&-\\\hline
      D$(5-\bar5)_1$&\tic&1$_0$,3$_1$&3$_1$&1$_0$\\\hline
      D$(5-\bar5)_3$&\tic&1$_0$,3$_3$&$3_3$&-\\\hline
      D$(3-\bar3)_0$&\tic&1$_0$&1$_0$&-\\\hline
      D$(1-\bar1)$&$\times$&&&\\\hline
    \end{tabular}
    \caption{Various inflatons and the cosmic strings to which they
    decay for a brane world built of sets of D$5_1$ and D$5_2$ branes.
    The cosmic string types allowed are determined by K-theoretic
    analysis of the non-BPS systems.  Since the Hubble size is
    greater than the compactification radii, the Kibble mechanism is
    capable of producing only defects localized in the three large
    spacial dimensions.  Cosmic strings can be thermally produced if
    unstable states are able to persist until reheating.
    }\label{inflatons}
\end{table}\end{center}
If a non-trivial 3-cycle is present in the orientifold model, 
a D5-brane wrapping such a cycle can appear as a domain wall, 
while a D3-brane wrapping it can appear as a monopole. 
However, such a 3-cycle will be in the (468) (or an equivalent) 
directions.  Since none of the Dp-\Dbar{p} pair that can generate
inflation wrap all these 3 directions, such defects are not produced.

Let us now elaborate on the various possibilities listed in
Table~\ref{inflatons} (below, a Dp-\Dbar{p} pair includes the case 
of a stack of Dp-branes separated from a stack of \Dbar{p}-branes):
\begin{itemize}
\item D9-\Dbar9 pair. In this case, the tachyon field is
  always present and the annihilation happens rapidly.  Also since the
  branes are coincident, there is no inflaton.
\item D1-\Dbar1 pair. Since they do not span the 3
  uncompactified dimensions, they do not provide the necessary
  inflation.  In the presence of inflation (generated by other pairs),
  a density of these D$1$-branes will be inflated away.
\item D(3-\Dbar3)$_0$ pair. They span the 3 uncompactified 
  dimensions and move towards each other inside the volume of the
  6 compactified dimensions during inflation. (The conservation of 
  the total zero RR charge prevents them from becoming parallel and 
  so BPS with respect to each other.) At the end of
  inflation, their collision heats the universe and yields D1$_0$-branes
  as vortex-like solitons. These D1$_0$-branes appear as cosmic
  strings. They form a gas of D1$_0$-branes (at all possible
  orientations in the 3-dimensional uncompactified space).  The
  D3$_0$-branes are unstable in the presence of the D5$_1$ and D5$_2$
  branes. It is possible that during inflation, the D3$_0$-brane can
  simply move towards a D5-brane and then dissolve into it. 
  The \Dbar3$_0$-brane can either hit the same D5-brane 
  ending inflation, producing
  D1$_0$-branes as cosmic strings, or it can collide with another
  D5-brane. This D5-brane shall no longer be BPS with respect to the 
  other D5-branes and more inflation may result from their
  interactions.  Towards the end of inflation these D5-branes collide
  with the BPS D5-branes.  D1$_0$-branes are expected to be produced as
  defects in this scenario.
\item D5$_1$-\Dbar5$_1$ pair.  This D5$_1$ brane is indistinguishable
  from the other D5$_1$-branes that are present.  They span the 3
  uncompactified dimensions and move towards each other inside the
  volume of the 4 compactified dimensions (\emph{i.e.}~(6789)) during 
  inflation. Towards
  the end of inflation, a tachyon field appears and its rolling
  produces D3$_1$ branes as cosmic strings. However such D3$_1$ branes 
  are unstable and eventually a tachyon field (an open string mode 
  between the D3- and the D5-branes) will emerge. Its rolling 
  signifies the dissolution of the D3-brane into the D5$_1$
  branes. Generically, by the time these D3-branes start dissolving,
  (re)heating of the universe should have taken place, so the tachyon
  rolling can thermally produce D1$_0$-branes as cosmic strings.
\item D5$_3$-\Dbar5$_3$ pair.  They may generate inflation 
  directly, and being mutually BPS with the D5$_1$ and D5$_2$ branes 
  shall not be subject to more complicated interactions.  After
  inflation, D3$_3$-branes as cosmic strings 
  will be produced.  Although they are not BPS with respect to the
  D5-branes, the interaction is repulsive (with $p=$5, $p^\prime$=3 
  and $a=$2 in Eq.(\ref{vpot})), so we expect them to move away from 
  the D5$_1$-branes in the (67) directions (to the anti-podal point) and 
  from the D5$_2$-branes in the (45) directions. This
  way, these D3$_3$-branes shall mostly survive and evolve into a cosmic
  string network. However, some of the D3$_3$-branes will scatter with 
  the D5-branes in the thermal bath. This may also result in the 
  production of some D1$_0$-branes as cosmic strings.
\item D7$_{1,3}$-\Dbar7$_{1,3}$ pair.  To provide the needed
  inflation, these pairs wrap 4 of the 6 compactified dimensions and
  move towards each other in the remaining 2 compactified dimensions
  during the inflationary epoch. Their collision heats the universe
  and yields D5$_{1,3}$-branes as cosmic strings.  The D5-branes that
  wrap only 2 of the 4 wrapped dimensions of the D7 branes may appear
  to simply span all 3 uncompactified dimensions.
  However, the production of these objects is severely suppressed
  since the Hubble size is much bigger than the typical
  compactification sizes. While the tachyon is falling down, the
  universe is still cold, so no thermal production is possible
  either. As a result, only D5$_{1,3}$-branes that appear as cosmic
  strings are produced.

  It is possible for the D5$_1$-branes to dissolve into magnetic flux
  on the D7-brane during inflation.  After the annihilation of the
  D7$_{1,3}$-\Dbar7$_{1,3}$ pair, this flux shall reemerge as 
  D5-branes, together with any additional D5-branes solitons as cosmic
  strings.
\item D7$_{1,2}$-\Dbar7$_{1,2}$ pair.  This case is similar to the
  above case, except both sets of D5-branes may dissolve into the D7
  pair during inflation.
\end{itemize}

We have considered only the IIB theory with two sets of D5-branes.
Under T-duality, the branes become D9-D5-branes, or D7-D3-branes in 
a IIB orientifold theory, with corresponding descriptions.
Generalizing the above analysis to the branes-at-angle scenario 
\cite{Garcia-Bellido:2001ky} should be interesting.
It is also possible to describe similar inflationary models with
cosmic strings in Type IIA theory, in which even dimensional branes 
are stable. In this case, one simply adds additional brane-anti-brane
pairs to the $\mathcal N=1$ spacetime supersymmetric IIA orientifold 
models \cite{csu}. It will be interesting to consider the brane 
inflationary scenario in M theory and the Horava-Witten model.
In general, we see that the brane inflationary scenario includes 
numerous possibilities, each with its own intriguing 
features and consequences.  

Although not necessary, we may consider the early universe starting as
a gas of branes (see for example \cite{Alexander:2000xv}).  The
presence of the orientifold planes fixes the total RR charge.  After
all but one pair of D-brane-anti-brane (that span the 3 large
dimensions) have annihilated, we end with an early universe that is
the starting point of the above discussion. In this picture, it is
hard to predict which set of D-brane-anti-brane should be last
standing.

\section{The spectrum of the cosmic strings}

The cosmic string tension $\mu$ is estimated for a number of 
brane inflationary scenarios \cite{Jones:2002cv,Sarangi:2002yt}.
The value $\mu$ is quite sensitive to the specific scenario.
Here we give an order of magnitude sketch.

For all brane separation smaller than the compactification size, 
the D-\Dbar potential is too steep for enough e-folding. 
When the brane and the anti-brane is far apart in the compactified 
volume, the images of the brane exert attractive forces on the 
anti-brane, so that at the anti-podal point the force is exactly 
zero. In the cubic compactification, this results in a potential
$V(\phi) = B - \hat \lambda \phi^4$,
where $\phi$ measures the distance from the anti-podal point 
\cite{Burgess:2001fx}.    
The density perturbation generated by the quantum fluctuation of 
the inflaton field is \cite{Burgess:2001fx,Jones:2002cv}
\be
\delta_H  \simeq  \frac{8}{5 \pi^2} \frac{N_e^{3/2}}{M_P r_{\perp}}
\ee
Using COBE's value $\delta_H \simeq 1.9 \times 10^{-5}$\cite{cobe},
\be
\label{perp}
M_pr_{\perp} \simeq 3 \times 10^{6}
\ee
This still leaves $M_s$ unfixed. To estimate $M_s$ and
the cosmic string tension $\mu$, let us consider a couple of scenarios.      
Consider D5$_1$-\Dbar5$_1$ brane inflation.
With $(M_sr_1)^2 \sim $30 and $r_2 = r_3 = r_{\perp}$, 
Eq.(\ref{mp}) and 
Eq.(\ref{perp}) then imply that $M_s \sim 10^{14}$ GeV.
If the cosmic strings are D$1$-branes, 
the cosmic string tension $\mu_1$ is simply 
the D$1$-brane tension $\tau_1$ : 
\be
        \mu_1 = \tau _1 = M_s^2/(2 \pi g_s)
\ee
This implies that $G \mu \simeq 6 \times 10^{-12}$.
Now the D1-brane may have discrete momenta in the 
compactified dimensions. These Kaluza-Klein modes give a 
spectrum of the cosmic string tension,
\be
     \mu \rightarrow \mu + e_1/r_1^2 + e_2/r_2^2 + e_3/r_3^2
\ee
where $e_i$ ($i=$1, 2, 3) are respectively the discrete eigenvalues 
of the Laplacians on the (45), (67) and (89) compactification cycles.
To get an order of magnitude estimate, we find that the lowest
excitation raises the tension by about a few percent.

For D7$_{1,2}$-\Dbar7$_{1,2}$ pair inflation, and 
$(M_sr_1)^2 \simeq (M_sr_2)^2 \sim $30, we have $r_3 = r_{\perp}$.
In this case, $M_s \sim 4 \times 10^{14}$ GeV,
with D5$_{1,2}$-branes as cosmic strings.
Noting that a Dp-brane has tension 
$\tau_p = M_s^{p+1}/(2 \pi)^p g_s$,
the tension of such cosmic strings is
\be
\mu_5 = (M_sr_1)^2(M_sr_2)^2 M_s^2/(2 \pi g_s)
\ee
This yields $G \mu \sim 10^{-8}$. 
This tension is bigger than that of D$1$-branes.
Depending on the particular inflationary scenario, this
value may vary by an order of magnitude. 
For D-\Dbar inflation, we have roughly \cite{Sarangi:2002yt}
\be
  10^{-7} \ge G \mu \ge 10^{-12}
\ee
Higher values of $G\mu$ are possible for
the branes-at-small-angle scenario. 

The interesting feature of this type of cosmic strings is that 
there is a spectrum of cosmic string tension.
The branes can wrap the compactified (4567)-dimensions more 
than once. This gives
\be 
\mu \sim n w \mu_5
\ee
where $n$ is the defect winding number (i.e., the vorticity) and $w$ 
is the wrapping number (i.e., the number of times it wraps the 
compactified volume) inside the brane, so $nw$ is equivalent to 
the number of cosmic strings. 
Moreover, there can be ``momentum'' (Kaluza-Klein) excitations 
of the branes propagating in these compactified directions. All these
result in quite an intricate spectrum of cosmic string tensions. 
For $n=w=1$, we have :
\be
\mu \sim \mu_5 \left(1 + \frac{p_1}{(M_sr_1)^2} +
\frac{p_2}{(M_sr_2)^2}\right)
 + \frac{e_3}{r_3^2}
\ee
where $p_1$ and $p_2$ are discrete ``momentum'' excitation modes 
depending on the geometry of the (45) and the (67) directions.
Using $(M_sr_1)^2 \sim (M_sr_2)^2 \sim $30, we see that each 
momentum excitation 
typically raises the cosmic string tension roughly by a few percent.

We see that the cosmic string tension can have a rich spectrum.
This is very different from the field theory case, where the cosmic 
string always appear with the same tension, up to the vorticity 
number $n$.

\section{Evolution of the cosmic string network}

To see the impact of the extra dimensions on the cosmic string network 
evolution, let us use the simple one-scale model for the evolution of the
cosmic string network \cite{Vilenkin:1981kz}.
The energy in the cosmic strings is much smaller than the energy 
in the radiation (or in the matter at later time). 
Let $L(t)$ be the characteristic length scale of the string network.
The energy density of the cosmic string network
is given by:
\be
\rho \simeq \frac{E}{L^3} \simeq \frac{\mu L}{L^3} 
\simeq \frac{\mu}{L^2}
\ee
where $E$ is the energy of the cosmic string network per 
characteristic volume.  String self-intersections typically break off
a loop, which then decays (\emph{e.g.}~via gravitational
waves). String intercommutations generate cusps and kinks, which also
decay rapidly.  So the change in energy is given by 
\be
\Delta E = \Delta E_{\text{expansion}} - \Delta E_{\text{interaction}}
\label{echange}
\ee
Now, the cosmic string energy in an expanding universe
$E= \rho V_0 a^3$ where the constant $V_0$ is the reference volume,
and $a(t)$ is the cosmic scale factor.
The number of interaction per unit volume per unit time
is $\lambda (v/L)/L^3$ where $v$ is a typical peculiar velocity
and $\lambda$ measures the probability of string intersections.
Assuming slow-moving strings (to simplify the analysis),
and substituting these quantities into Eq.(\ref{echange}),
we obtain the equation governing the evolution of the energy density:
\be
\label{netden}
\dot\rho = -2\frac{\dot a}{a}\rho - \lambda\frac{\rho}{L}
\ee
Here, $H={\dot a}/a$ is the Hubble constant. 
Substituting the ansatz for $L\left(t\right)=\gamma\left(t\right)t$
in Eq.(\ref{netden}),
we obtain the following equation for
$\gamma\left(t\right)$ during the radiation dominated era:
\be
\dot\gamma=-\frac{1}{2t}\left(\gamma-\lambda \right)
\ee
This equation has a stable fixed point at $\gamma(t) =\lambda$. 
We see from this 
solution that the characteristic length scale of the string network 
tends asymptotically towards the horizon size, $L\simeq \lambda t$. 

As a check, we see that in the absence of string interactions 
(that is, $\lambda=0$), $\gamma \sim \sqrt{t}$ so $\rho \sim a^{-2}$, 
as expected. In the presence of cosmic string interactions (that is, 
$\lambda \ne 0$),
the asymptotic (late time) energy density of the cosmic string is 
given by 
\be
\rho=\frac{\mu}{\lambda^2t^2} \sim 
\left\{  \begin{array}{ll} 
{\mu}/{\left(\lambda^2a^4\right)} & \text{radiation dominated era} \\
{\mu}/{\left(\lambda^2a^3\right)} & \text{matter dominated era}
\end{array}\right.
\ee
Suppose $\lambda_0$ is the interaction strength for field theory models. 
The cosmic strings live in $4+\dper$ dimensions and are localized in the
$\dper$ compact dimensions. The effect of the extra dimensions is to 
reduce the collision (self-intersection) probability of the
cosmic strings. The simplest way to model this effect is to change 
the efficiency with which loops are formed by the long cosmic strings, 
so $\lambda < \lambda_0$. So the number density of the scaling cosmic 
string network is enhanced by a factor of 
$$(\lambda_0/\lambda)^2 \ge 1$$ Generically, we expect this 
to be a large enhancement. 
Since the extra dimensions are stabilized, $\rho$ still 
scales like radiation during the radiation-dominated epoch
(and it scales like matter during the matter-dominated epoch).
The resulting cosmic string network then yields a scaling energy 
density $\rho$
\be 
\frac{\rho}{\rho_r} \simeq \Gamma G \mu 
\simeq \left(\frac{\lambda_0}{\lambda}\right)^2 \beta G \mu 
\ee
where $\rho_r$ is the energy density of radiation 
during the radiation-dominated epoch (or of matter during the 
matter-dominated epoch).
In field theory models, $\Gamma =\beta$.
Numerical simulations \cite{Bennett:1990uz} give $\beta \sim 6$. 
%So the effect on the cosmic string network can be very dramatic.

Let us give an order-of-magnitude estimate of
the effect of the extra dimensions on the cosmic string collision
probability, namely the ratio $\lambda_0/\lambda$.

Consider two points of two different cosmic strings or of the same 
cosmic string that coincide in the 4-dimensional spacetime. In 
4-dimensional field theory, they are touching. The probability 
of this happening is dictated by $\lambda_0$. In the brane world, 
they may still be separated in the extra dimensions. We like to 
estimate the likeliness of them actually touching (which then allows 
intercommuting or the pinching off of a loop).

Consider the compact directions where these two points (of cosmic 
strings) appear as points (that is, they are not wrapping these 
compact directions).
In the case of D5$_3$-\Dbar5$_3$ pair inflation, the 
repulsive force from the D5$_1$-branes will push 
the D3$_3$-branes into a corner in the (45) directions, while
the repulsive force from the D5$_2$-branes will push the 
D3$_3$-branes into a corner in the (67) directions. 
As a result, all the D3$_3$-branes end up at a corner in the
(4567) directions. In this case, the extra dimensions should 
have little or no effect on their interaction, that is
\be
\lambda_0/\lambda \sim 1 \quad \quad \text{for} \quad 
\text{D3}_3\text{-branes}
\ee
In other scenarios, they are free to roam in the compact directions.
If two cosmic strings coincide in the 4-dimensional spacetime, and in
the compactified directions in which they are pointlike they are
separated by a distance comparable to the superstring scale 
$1/M_s$, a tachyon field appears and the rolling of this
tachyon field has a time scale around the superstring scale. 
So we expect them to interact. Consider the scenario where the
D$5_{1,2}$-branes are cosmic strings. If they are randomly placed,
the likeliness of them coming within 
that distance in the compact (89) directions is given by
$\lambda_0/\lambda \simeq (M_sr_3)^2$.
Now let us take the cosmic string interaction into account.
Since the cosmic string appears  
as points and interact via an attractive Coulomb type potential 
in the extra dimensions (which becomes important only
when the separation between them is relatively small). Let us get 
an estimate of this enhancement of the probability of interaction.
The scattering cross-section of the two string points interaction 
in transverse dimensions via an attractive potential
$V\left(r\right)=-A/r^{\dper-2}$ is given by:
\be
\sigma = \Omega_{\dper-1}r_{\text{capture}}^{\dper-1}
\ee
where $\Omega_{\dper-1}$ is the volume of the unit
$(\dper-1)$-sphere. The capture radius $r_{\text{capture}}$ is
comparable to the superstring scale $1/M_s$, so the likelihood 
of two string points within that distance becomes
\be
\frac{\lambda_0}{\lambda} \simeq \frac{r_3}{r_{\text{capture}}}
\simeq  M_sr_3 \sim 10
\ee
Note that, generically, larger $G\mu$ 
gives less enhancement. The reason is :
the total volume of the compactified dimensions is fixed by 
the value of $G$; larger tension comes from brane-wrapping over 
larger compactified volume, which implies smaller
volume for the cosmic strings to avoid each other, so smaller
$\Gamma$. This means $\Gamma G \mu$ (the cosmic string density) is 
relatively insensitive compared to either $\mu$ or $\Gamma$ alone. 
If observations give a bound $\rho/\rho_r < 10^{-5}$, 
the values of $G\mu$ appearing in the branes-at-small-angle 
scenario may seem too large. However, the production of cosmic 
strings in this scenario are more localized around the brane 
intersection, implying a smaller production as well as a 
smaller enhancement in $\Gamma$.
Clearly, a careful estimate of $\lambda_0/\lambda$ 
and $\Gamma$ in that case will be very important. 

The important message here is that the comic string network  
continues to have a scaling solution, and the 
enhancement in its energy density due to the extra dimensions
can be very large. For a given $\mu$, this will yield a very 
different cosmic string energy density than that in the field 
theory case. Measuring $\mu$ and $\rho$ separately will be valuable. 

Dvali and Vilenkin also noted that the presence of compactified 
dimensions can substantially increase the number density of the cosmic 
string network.
We thank Louis Leblond, Levon Pogosian, Sash Sarangi, Gary Shiu, 
Alex Vilenkin and Ira Wasserman for valuable discussions.
This research is partially supported by the National Science
Foundation under Grant No. PHY-0098631.


\begin{thebibliography}{10}

\bibitem{cobe} G.~F.~Smoot \emph{et.~al.}, 
{\em Astrophy.~J.} {\bf 396} (1992) L1; \\
C.~L.~Bennett \emph{et.~al.}, {\em Astrophy.~J.} {\bf 464} (1996)
L1, {\tt astro-ph/9601067}.

\bibitem{new} A.~T.~Lee \emph{et.~al.} (MAXIMA-1), {\em Astrophys.~J.} 
{\bf 561} (2001) L1, {\tt astro-ph/0104459};\\
C.~B.~Netterfield \emph{et.~al.} (BOOMERANG), {\em Astrophys.~J.} 
{\bf 571} (2002) 604 
{\tt astro-ph/0104460}; \\
C.~Pryke, \emph{et.~al.} (DASI), {\em Astrophys.~J.} {\bf 568} (2002) 46 
{\tt astro-ph/0104490}; \\
H.~V.~Peiris, \emph{et.~al.} (WMAP), {\tt astro-ph/0302225}.

\bibitem{guth}A.~H.~Guth, {\em Phys. Rev.} {\bf D23} (1981) 347; \\
A.~D.~Linde, {\em Phys. Lett.} {\bf B108} (1982) 389; \\
A.~Albrecht and P.~J.~Steinhardt,
{\em Phys. Rev. Lett.}  {\bf 48} (1982) 1220.

\bibitem{Sen:1999mg}
A.~Sen, %{\it Non-{BPS} states and branes in string theory},
% Report: MRI-PHY/P990411 
{\tt hep-th/9904207}.

\bibitem{Dvali:1998pa}
G.R.~Dvali and S.-H.~H.~Tye, %{\it Brane inflation},  
{\em Phys. Lett.} {\bf B450} (1999) 72, {\tt hep-ph/9812483}.

\bibitem{Burgess:2001fx}
C.~P.~Burgess, M.~Majumdar, D.~Nolte, F.~Quevedo, G.~Rajesh, and R.~Zhang,
%{\it The inflationary brane-antibrane universe},  
{\em JHEP} {\bf 07} (2001) 047, {\tt hep-th/0105204}.

\bibitem{Garcia-Bellido:2001ky}
J.~Garcia-Bellido, R.~Rabad\'an, and F.~Zamora, 
%{\it Inflationary scenarios from branes at angles},  
{\em JHEP} {\bf 01} (2002) 036, {\tt hep-th/0112147}.

\bibitem{Jones:2002cv}
N.~Jones, H.~Stoica, and S.-H.~H.~Tye, 
%{\it Brane interaction as the origin of inflation},  
{\em JHEP} {\bf 07} (2002) 051, {\tt hep-th/0203163}.

\bibitem{Quevedo:2002xw}
G.~Dvali, Q.~Shafi and S.~Solganik, {\tt hep-th/0105203}; \\
G.~Shiu and S.-H.~H.~Tye, {\em Phys. Lett.} {\bf B516} (2001) 421, 
{\tt hep-th/0106274}; \\
C.P.~Burgess, P.~Martineau, F.~Quevedo, G.~Rajesh and R.-J. Zhang, 
JHEP 0203 (2002) 052, {\tt hep-th/0111025}; \\
R.~Blumenhagen, B.~Kors, D.~Lust and T.~Ott, 
{\em Nucl. Phys.} {\bf B641} (2002) 235, {\tt hep-th/0202124}; \\
K.~Dasgupta, C.~Herdeiro, S.~Hirano, and R.~Kallosh, 
%{\it {D3/D7} inflationary model and {M}-theory},  
{\em Phys. Rev.} {\bf D65} (2002) 126002,
{\tt hep-th/0203019};\\
M.~Gomez-Reino and I.~Zavala, 
%{\it Recombination of intersecting {D}-branes and cosmological inflation},  
{\em JHEP} {\bf 09} (2002) 020, {\tt hep-th/0207278}; \\
F.~Quevedo, 
%{\it Lectures on string / brane cosmology},  
{\em Class. Quant. Grav.} {\bf 19} (2002) 5721, {\tt hep-th/0210292};\\
R.~Brandenberger, G.~Geshnizjani and S. Watson, {\tt hep-th/0302222}.
\bibitem{stw} G.~Shiu, S.-H.~H.~Tye and I.~Wasserman, {\tt hep-th/0207119};\\
J.~M.~Cline, H.~Firouzjahi and P.~Martineau, JHEP {\bf 0211} (2002) 041 
{\tt hep-th/0207156}; \\
G.~Felder, J.~Garcia-Bellido, P.~B.~Greene, L.~Kofman,
A.~Linde and I.~Tkachev, Phy. Rev. Lett. {\bf 87} (2001) 011601; 
{\tt hep-ph/0012142}\\
G.~Felder, L.~Kofman and A.~Starobinsky, JHEP 0209 (2002) 026, 
{\tt hep-th/0208019}.

\bibitem{Alexander:2001ks}
S.~H.~S.~Alexander, 
%{\it Inflation from d - anti-d brane annihilation},  
{\em Phys. Rev.} {\bf D65} (2002) 023507, {\tt hep-th/0105032};\\
M.~Majumdar and A.-C.~Davis, 
%{\it Cosmological creation of {D}-branes and anti-{D}-branes},  
{\em JHEP} {\bf 03} (2002) 056, {\tt hep-th/0202148}.

\bibitem{Sarangi:2002yt}
S.~Sarangi and S.-H.~H.~Tye, 
%{\it Cosmic string production towards the end of brane inflation},  
{\em Phys. Lett.} {\bf B536} (2002) 185, {\tt hep-th/0204074}.

\bibitem{Gimon:1996rq}
E.~G.~Gimon and J.~Polchinski, 
%{\it Consistency conditions for orientifolds and {D}-manifolds},  
{\em Phys. Rev.} {\bf D54} (1996) 1667, {\tt hep-th/9601038};\\
A.~Dabholkar and J.~Park, 
%{\it An orientifold of type-{IIB} theory on {K}3},
{\em Nucl. Phys.} {\bf B472} (1996) 207, {\tt hep-th/9602030};\\
E.~G.~Gimon and C.~V.~Johnson, 
%{\it {K}3 orientifolds},  
{\em Nucl. Phys.} {\bf B477} (1996) 715, {\tt hep-th/9604129};\\
C.~Angelantonj, M.~Bianchi, G.~Pradisi, A.~Sagnotti and Ya.S.~Stanev, 
{\em Phys. lett.} {\bf B385} (1996), {\tt hep-th/9606169}; \\
Z.~Kakushadze, G.~Shiu, and S.-H.~H.~Tye, 
%{\it Type {IIB} orientifolds with {NS-NS} antisymmetric tensor backgrounds},  
{\em Phys. Rev.} {\bf D58} (1998) 086001, {\tt hep-th/9803141};\\
G.~Aldazabal, A.~Font, L.~E.~Ibanez, and G.~Violero, 
%{\it {D} = 4, {N} = 1, type {IIB} orientifolds},  
{\em Nucl. Phys.} {\bf B536} (1998) 29, {\tt hep-th/9804026} \\
G.~Shiu and S.-H.~H.~Tye,
{\em Phys. Rev.} {\bf D58} (1998) 106007, {\tt hep-th/9805157}; \\
Z.~Kakushadze, {\em Int. J. Mod. Phys.} {\bf A15} (2000) 3113,
{\tt hep-th/0001212}.

\bibitem{Berkooz:1996km}
M.~Berkooz, M.~R.~Douglas, and R.~G.~Leigh, 
%{\it Branes intersecting at angles},  
{\em Nucl. Phys.} {\bf B480} (1996) 265, {\tt hep-th/9606139}.

\bibitem{Lifschytz:1996iq}
G.~Lifschytz, 
%{\it Comparing {D}-branes to black-branes},  
{\em Phys. Lett.} {\bf B388} (1996) 720, {\tt hep-th/9604156};\\
H.~Arfaei and M.~M.~Sheikh~Jabbari, 
%{\it Different {D}-brane interactions},
{\em Phys. Lett.} {\bf B394} (1997) 288, {\tt hep-th/9608167}.

\bibitem{Gava:1997jt}
E.~Gava, K.~S.~Narain, and M.~H.~Sarmadi, 
%{\it On the bound states of p- and (p+2)-branes},  
{\em Nucl. Phys.} {\bf B504} (1997) 214, {\tt hep-th/9704006}.

\bibitem{csu}
M.~Cvetic, G.~Shiu and A.~M.~Uranga, 
{\em Nucl. Phys.} {\bf B615} (2001) 3, {\tt hep-th/0107166}.

\bibitem{Kakushadze:1998wp}
Z.~Kakushadze and S.-H.~H.~Tye, 
%{\it Brane world},  
{\em Nucl. Phys.} {\bf B548} (1999) 180, {\tt hep-th/9809147}.

\bibitem{Witten:1998cd}
A.~Sen, {\em JHEP} {\bf 9808} (1998) 012, {\tt hep-th/9805170}; \\
E.~Witten, 
%{\it {D}-branes and {K}-theory},  
{\em JHEP} {\bf 12} (1998) 019, {\tt hep-th/9810188}.

\bibitem{Alexander:2000xv}
S.~Alexander, R.~H.~Brandenberger and D.~Easson, 
%{\it Brane gases in the early universe},  
{\em Phys. Rev.} {\bf D62} (2000) 103509, {\tt hep-th/0005212}.

\bibitem{Vilenkin:1981kz}
A.~Vilenkin, 
%{\it Cosmic strings},  
{\em Phys. Rev.} {\bf D24} (1981) 2082.

\bibitem{Bennett:1990uz}
A.~Albrecht and N.~Turok, 
{\em Phys. Rev. Lett.} {\bf 54} (1985) 1868;\\
D.~P.~Bennett and F.~R.~Bouchet, 
%{\it High resolution simulations of cosmic string evolution: 
%Part 1. numerics and long string evolution.},  
{\em Phys. Rev.} {\bf D41} (1990) 2408 ; \\
B.~Allen and E.~P.~S.~Shellard, 
{\em Phys. Rev. Lett.} {\bf 64} (1990) 119.


\end{thebibliography}
\end{document}